\documentclass[sigconf]{acmart}
\usepackage{todonotes}
\usepackage{subcaption}
\usepackage{makecell}

\AtBeginDocument{%
  }

\copyrightyear{2026}
\acmYear{2026}
\setcopyright{cc}
\setcctype{by}
\acmConference[WebSci Companion '26]{18th ACM Web Science Conference}{May 26--29, 2026}{Braunschweig, Germany}
\acmBooktitle{18th ACM Web Science Conference (WebSci Companion '26), May 26--29, 2026, Braunschweig, Germany}
\acmDOI{10.1145/3795513.3807439}
\acmISBN{979-8-4007-2492-3/2026/05}

\begin{document}

\title{Citation Farming on ResearchGate: Blatant and Effective}

\author{Cenk Erdoğan}\email{c.erdogan@student.maastrichtuniveristy.nl}
\authornote{All authors contributed equally to this research.}
\author{Bennett Daniel}\email{bennett.daniel@student.maastrichtuniveristy.nl}
\authornotemark[1]
\author{Benedikt Wotka}\email{b.wotka@student.maastrichtuniveristy.nl}
\authornotemark[1]
\author{Ashish Sai}
\author{Adriana Iamnitchi}
\affiliation{%
  \institution{Maastricht University}
  \city{Maastricht}
  \country{Netherlands}  
}


\begin{abstract}

We investigate platform-native citation farming on ResearchGate by analyzing almost 3000 papers uploaded by five suspected boosting-service provider accounts. From the uploaded papers and associated metadata, we construct both paper-level and author-level citation networks. We introduce an interpretable structural signal for coordinated boosting, \emph{equal references groups}: clusters of papers with equal reference lists. We find that many papers from our collection exhibit this motif, that is, they disproportionately cite a small set of authors, consistent with coordinated or automated boosting rather than independent scholarly practice. Finally, we show that for some authors in our dataset a substantial share of their citations can be attributed to these suspicious groups. A different citation network was used to validate the rareness of such motifs in legitimate scientific work.

\end{abstract}

\maketitle

\section{Introduction}
The internet is being flooded with content produced by large language models (LLMs), a development that has already reached the domain of scientific publications \cite{liang2024mapping}. This trend is facilitated by platforms such as ResearchGate, which allow paper uploads without verification or peer review, thereby enabling the unregulated spread of AI-generated material. In addition to polluting scientific knowledge and discrediting scientific practices, some of these LLM-generated papers are used to increase the visibility of some beneficiary authors by citing their work and thus boosting their citation counts. 

Manipulated citations are highly problematic: they distort the value of scholar recognition and render irrelevant traditional academic productivity metrics. In addition, via misatribution of such LLM-generated papers, they may tarnish the reputation of authentic researchers. The propagation of scientific misinformation is further helped by the uncritical citation of these illegitimate articles in otherwise legitimate publications.  

Previous research on citation behavior and manipulation focused on multiple aspects. First, research identified structural biases affecting citation rates~\cite{norgaard2026predictors} that, while not considered purposely malicious, are not related to content value. 

Second, various studies focused on identifying suspicious citation patterns via network-based methodologies. For example, Avros et al.~\cite{Avros2023} introduced a perturbation and embedding approach to randomly remove edges from a citation network and attempt to reconstruct them. Edges that fail to reappear reliably are flagged as suspicious. Tests show that many citation edges are structurally unstable, suggesting possible manipulation.
Liu et al.~\cite{Liu2022} use a deep‑graph model that combines network topology with textual/semantic citation context, and a ``Citation Purpose” algorithm, GLAD, to distinguish plausible from anomalous citations. Evaluated on a simulated anomalous-citation dataset, GLAD significantly outperforms baseline link‑prediction methods, demonstrating that content‑aware network analysis improves detection of potentially fake citations.

Third, few studies present empirical evidence of citation manipulation. Ibrahim et al.~\cite{ibrahim_citation_2025} demonstrated that citation‑boosting services are effective in manipulating widely used academic performance metrics reported by Google Scholar. 
Wren at al.~\cite{Wren2022}, in a study of more than 20,000 authors from PubMed, found that the distribution of non‑self‑citations of a given author from a single citing paper is inversely proportional to their rank. The study estimates that up to $\approx$16\% of the authors in their dataset may have engaged in reference list manipulation to some degree. More relevant to our work, Kirilova and Zoepfl~\cite{KIRILOVA2025101604} showed that academic evaluation metrics are susceptible to manipulation, including via the production of fraudulent papers designed to inflate citations. 

Our work is motivated by the experience of one of the authors, who discovered on Google Scholar an article falsely attributed to them. While the topic aligned with their genuine area of expertise, the listed co-authors did not exist at the stated institutions, the paper was strangely formatted, surprisingly coherent in form and nonsensical in substance. Due to this unusual discovery, our work differentiates itself from previous articles in various ways. First, we are able to pursue an actor-seeded, forensic discovery workflow (starting from suspected providers). Instead of scanning an entire corpus for anomalies, we introduce a seed-based investigative approach: we begin with suspected boosting-service provider accounts and expand to the surrounding papers/authors/citations. This is a distinct methodological stance compared to global anomaly detection, which is pursued in other work~\cite{ibrahim_citation_2025}. Second, due to this ground truth dataset of fraudulent papers, we can analyze the methodology employed by these citation boosting services with more accuracy. Moreover, in this clean dataset we can also detect who the beneficiaries are and can estimate the impact of these fraudulent citations on their bibliometrics.  

Our contributions are the following: (a)~a platform-native study of citation farming on ResearchGate; (b)~an actor-seeded forensic workflow that starts from suspected boosting-provider accounts and expands to the beneficiary papers/authors; (c)~an interpretable structural signal for detection via equal reference groups (citation motifs); and (d)~the dataset collected and used in this study, publicly available at~\cite{rg}.




\section{Data Collection}
A web search on the paper claming to have been co-authored by one of the authors directed the search to a suspected boosting service provider account (SSPA) on ResearchGate. The supposed author does not exist outside RG, despite having published hundreds of papers and co-authored work across several fields.
A broader search revealed four additional accounts with similar characteristics. These accounts also shared co-authored publications with the original profile. In total, five authors (SSPA) with similar suspicious profiles have been detected with seemingly LLM-generated papers uploaded on ResearchGate.

We started our research with the list of these five initial suspected boosting service provider accounts, and the papers they had uploaded to ResearchGate. On RG, uploaded PDF files are accompanied by platform-generated metadata including title, author list, affiliations, publication venue, and reference lists, which are extracted automatically and displayed separately from the document itself. Each paper and author is uniquely identified through persistent profile links.

Because the formatting and structure of uploaded paper files varied widely, automated extraction of bibliographic references was unreliable. Instead, we relied on the metadata provided by ResearchGate by scraping the platform. This specifically helped with the identification of existing and non-existent references, as well as uniquely identifying all authors and papers via their profile links.
We first collected all papers uploaded by the five SSPAs from their ResearchGate profile pages and scraped metadata from each paper’s dedicated page. For authors with RG profiles, we additionally retrieved profile information and publication metrics, noting that most LLM-generated authors did not maintain profiles.


In total, we were able to collect the following dataset: we recorded a set of 2,988 seed papers that cited a total of 12,786 articles. Our seed papers were collectively cited 6,221 times. We collect a total of 22,462 authors across all papers and references, 7,124 of whom have ResearchGate profiles. We additionally collected profile information for 4,048 authors of cited papers.
Due to dynamically loaded content, we captured approximately one quarter of references and citations. This may have excluded some additional suspicious references, which we consider a limitation.
With the collected data, we created two directed citation networks, one with authors as nodes, and the other with papers as nodes. The resulting dataset is publicly available~\cite{rg}.




\section{Fake Authors' Patterns of Publication}

The SSPAs are currently publishing several papers per month, most of which are preprints with seemingly non-existent\footnote{By “non-existent,” we refer to co-authors whose names and institutional affiliations could not be independently verified using publicly available information.} co-authors. 

\begin{figure}[htbp]
    \centering
    \includegraphics[width=0.9\linewidth]{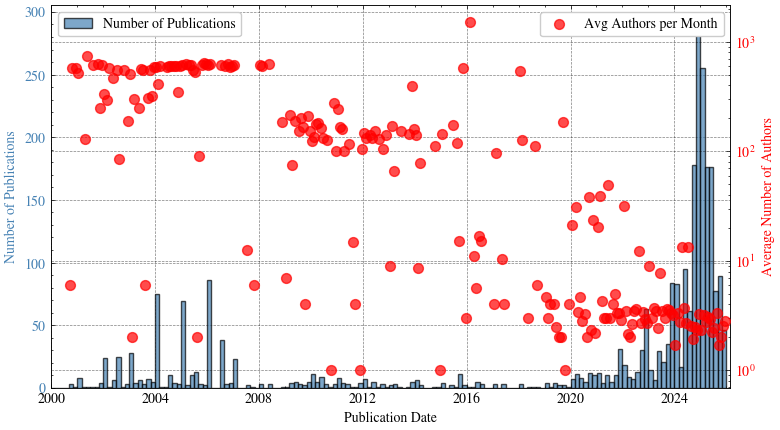}
    \caption{Timeline of illegitimate account publications with author count (binned in buckets of three months, only publications since 2000).}
    \label{fig:monthly_pubs_authors}
\end{figure}

We found two different types of papers. Papers published before $\approx$2022 have many more authors and are seemingly legitimate. One such example is~\cite{600_aubert2007measurement}. We suspect that these papers have been re-uploaded by the SSPAs to claim co-authorship and therefore garner citations and co-authorships from legitimate research. While these papers were published several years ago, they seem to only haven been uploaded recently.

Since 2022 the SSPAs have begun publishing several papers per month. Papers would then, for example, be authored by \textit{SSPA-1}, John Smith and Jane Doe. These papers also had higher rates of not being published anywhere else. 

Based on qualitative sampling, we identified additional suspicious patterns. The more recent papers often contained little or no author information, and uploaded filenames were nearly identical, differing only by sequential numbering. The text typically consisted of loosely connected bullet points or very short paragraphs, with minimal formatting and poorly structured references; reference lists were either extremely short or dominated by a small set of authors. Some papers appeared superficially well produced and included figures or tables, despite the underlying text being largely nonsensical. In a few cases, papers were published in existing journals that appeared to accept work with little regard for scientific quality.
These observations have not yet been systematically quantified and may reflect differences in how the papers are generated or on the requirements of the customers.

\section{Detecting Authors with Boosted Citations}

To obtain accurate results, we build on the approach introduced by Ibrahim et al. \cite{ibrahim_citation_2025} by focusing on the analysis of the paper–citation graph and the detection of citation motifs.
In these motifs, many papers by supposedly different authors shared an identical set of references. These references also often contained many publications by the same author. We identify groups of authors that consist of \textit{citer} and \textit{cited} and use them to capture such motifs. We define an \textbf{equal references group} a set of papers whose reference lists are equal. We use these papers with identical reference lists to build a similarity network. In this network, each node represents a paper, and two papers are connected if they have equal references. A group is defined as a maximal clique: a set of papers such that every paper in the set is similar to every other paper in the set, and no additional papers can be added without breaking this similarity. These groups indicate very unnatural behaviour and imply a relationship between group members. 

\begin{figure}[ht]
    \centering
    \begin{subfigure}[t]{0.4\linewidth}
        \centering
        \includegraphics[width=\linewidth]{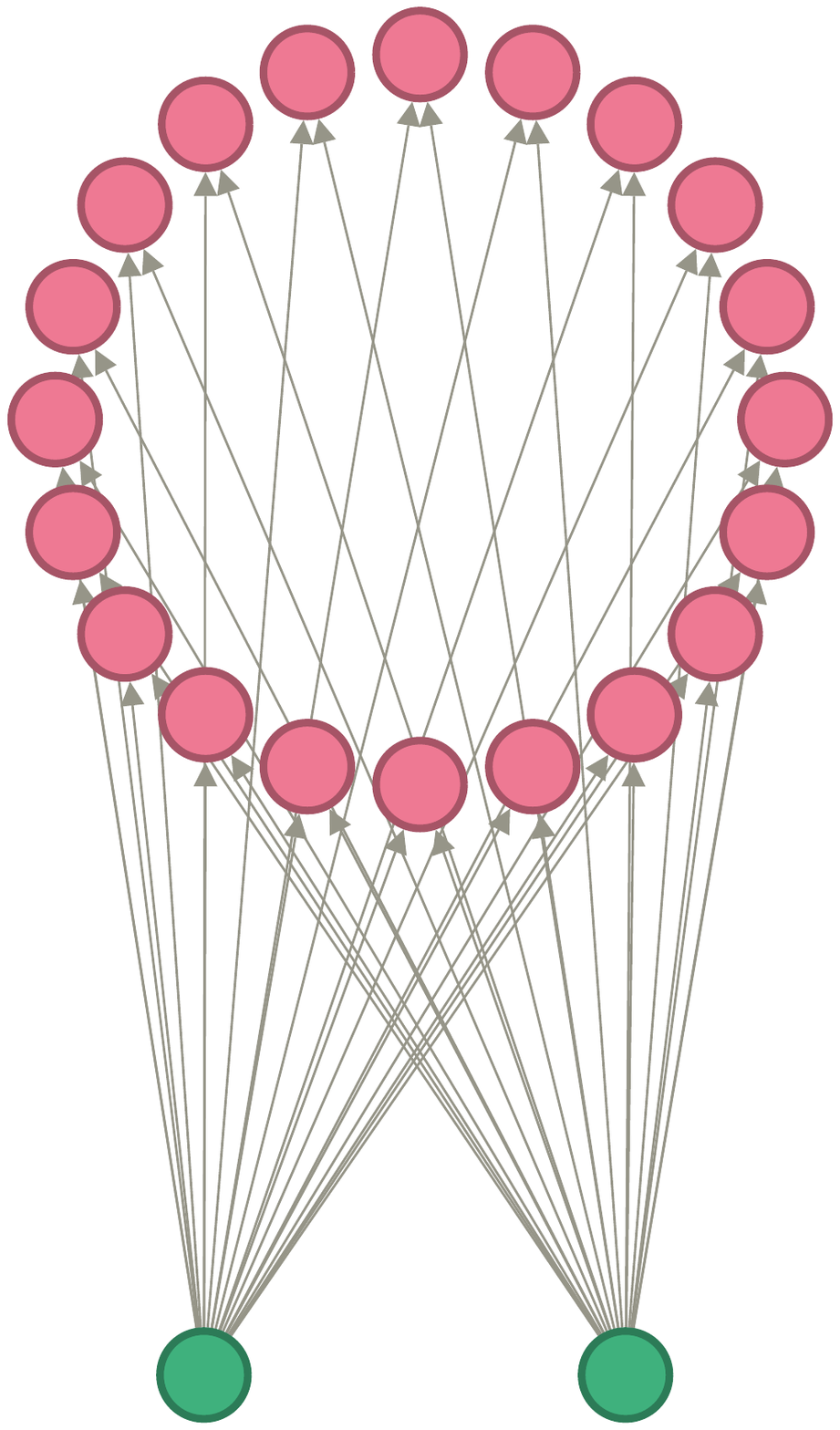}
        \label{fig:unnatural1}
    \end{subfigure}
    \hfill
    \begin{subfigure}[t]{0.4\linewidth}
        \centering
        \includegraphics[width=\linewidth]{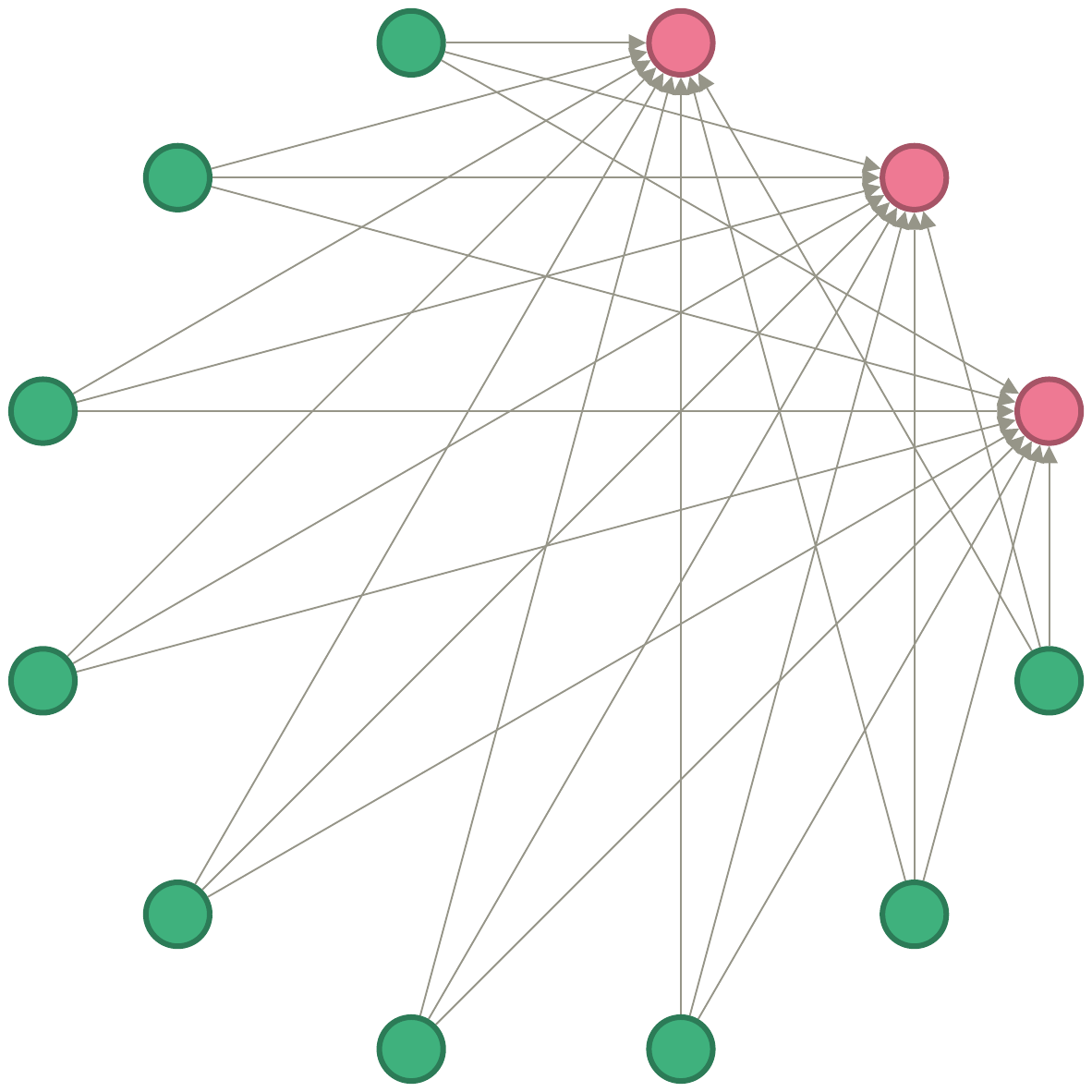}
        \label{fig:unnatural2}
    \end{subfigure}
    \caption{Two equal references groups. Green nodes denote citing paper who have identical outgoing citation sets and do not cite any papers outside the depicted group. Red nodes represent cited papers. }
    \label{fig:unnatural_both}
\end{figure}

The groups we identified provide a framework for uncovering accounts that may be operated by the same underlying actor, as well as publications that receive disproportionate benefits from these atypical motifs. In genuine academic work, even closely related work in the same field substantially differentiate in their reference list. The groups we identify unravel patterns that should not occur naturally, as can be seen in two examples in Fig. \ref{fig:unnatural_both}.
These kind of motifs strongly suggests that the citing papers (coloured in green) have been produced with an ulterior motive in mind. These structures seem to imply that at least one involved actor intends to benefit from it, especially when credit is consistently directed toward the same set of authors by repeated citation structures. The inclusion of other authors in these reference lists are used to hide the underlying manipulation. Therefore, authors can appear merely to make the pattern appear more natural.

In total, we identified 240 citation groups, each consisting of two or more authors whose papers reference exactly the same set of papers. As shown in Figure \ref{fig:unnatural_both}, each group member acts either as a \textit{citer} or as \textit{cited}.

Because citation inflation operates at the author level rather than the paper level, determining which authors appear most frequently across groups is a big indicator in identifying the beneficiaries.
Another important metric is how many distinct works of a cited author appear in any groups, hinting at their work being selected to appear instead of randomly appearing because of their relevance in the research field.

\begin{figure}
    \centering
    \includegraphics[width=0.9\linewidth]{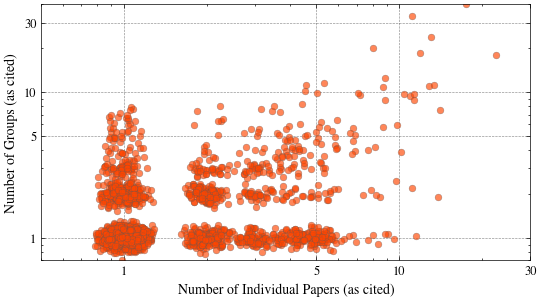}
    \caption{Scatter plot showing the relationship between the number of distinct papers cited per author (x-axis) and the number of groups in which the author appears as cited (y-axis)}
    \label{fig:scatter_author_papers}
\end{figure}

Figure \ref{fig:scatter_author_papers} shows the relationship between the number of cited papers of an author and the total number of groups in which that author appears as cited. Most authors are in the lower-left region, few papers are cited and appearance in few groups. A few authors are in the upper-right region, meaning many publications are cited repeatedly across many groups. These outliers help us identify the major beneficiaries of artificial citation efforts. However, many individuals who purchase small amounts of citations may not appear as statistical extremes. Anyone appearing in these groups could be buying citations. Presence alone does not prove involvement.


To further validate whether beneficiaries which were cited in the detected motifs had meaningfully improved their metrics by appearing in the motifs, we compared the number of times they appeared in the motifs to their total citations on ResearchGate. If they were already a well-known author which was simply cited several times due to valuable contributions to their field, the share of citations they get from inside the motifs should be relatively small. Some anonymised examples can be found in Table \ref{tab:sus_authors}.
\begin{table}[h!]
\caption{Top beneficiaries by share of motif citations, including highly cited outliers}
\label{tab:sus_authors}
\centering
\begin{tabular}{c|r|r|r}
\makecell{\textbf{Author}\\\textbf{Hash}} &
\makecell{\textbf{Total}\\\textbf{Citations}} &
\makecell{\textbf{Citations from}\\\textbf{within motifs}} &
\makecell{\textbf{Share of}\\\textbf{motif citations}} \\
\hline
d0ff66 & 27 & 22 & 81\% \\
e06c8a & 90 & 52 & 58\% \\
7f056b & 42 & 24 & 57\% \\
594ad6 & 436 & 224 & 51\% \\
bf80bb & 110 & 49 & 45\% \\
\end{tabular}
\end{table}

The identified motifs are based solely on the data we scraped. For the author with the hash '594ad6', we found an even larger number of citations from suspicious papers which have not yet been scraped.

\section{Comparison to existing citation networks}
\begin{figure}[h]
    \centering
    \includegraphics[width=0.9\linewidth]{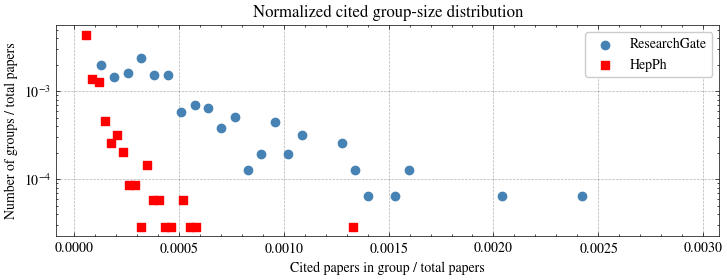}
    \caption{Comparison of numbers of groups and groups size between HepPh and the ResearchGate dataset}
    \label{fig:comp_size_number}
\end{figure}

To further underline just how unusual the motifs we found were, we have ran the same motif detection on the HepPh citation network., introduced in \cite{gehrke2003overview}. HepPh contains almost 35,000 papers from the field of high energy physics phenomenology. The HepPh dataset is limited in the sense that citations to papers which are not in the network are not recorded. Two papers at the edge of the network which cite eight different papers outside the network and the same two papers from inside the network would appear in the same motif despite having distinct reference lists.
This leads to a large number of motifs with a small number of overlapping. While HepPh has  more motifs (309 vs 240), the majority of the motifs are either with few overlapping cited or few citing papers. HepPh has significantly fewer groups of every size except for very small motifs as can be seen in Figure~\ref{fig:comp_size_number}. Numbers are normalized by number of papers.

\section{Summary and Discussions}
This paper has shown that some service providers likely take advantage of platforms such as ResearchGate to boost the citation counts of existing academic authors (beneficiaries) as part of a business transaction. While this phenomenon has been suspected before, our investigation focuses on ResearchGate data starting from a small set of suspected boosting service provider accounts (SSPAs). We collected data on 2,988 papers uploaded by five SSPAs and used network analysis techniques to uncover unusual network motifs. These motifs capture patterns in which several papers share their reference lists, implying either heavy plagiarism or a coordinated effort to increase the citation counts of specific beneficiaries. Most notably, we identify 240 distinct groups, which contain 2 or more papers who are characterized by an exact match in their reference lists.
We also compared the found motifs to the HepPh dataset \cite{gehrke2003overview} to further show how unusual the appearnce of these motifs is.

These motifs can be used both to identify anomalies in the citation network (and thus, the service providers) and to highlight the beneficiaries who benefit from citation boosting. We identified several beneficiaries who obtain a substantial share of their total citations from these suspicious papers. These authors were cited by multiple papers that shared most of their references and received a large proportion of their ResearchGate citations from these papers. This further confirms earlier concerns that raw citation counts are an insufficient and fragile measure of academic performance.
We also observed tactics that SSPAs use to boost their credibility, such as fraudulently claiming co-authorship with authentic scientists and manipulating the format and appearance of publications. 

Previous work~\cite{ibrahim_citation_2025} suggested that citation boosting has become a business model. We encountered several advertisements for such services and contacted the providers to learn more about their practices. Because all the accounts used to upload fraudulent papers had invented names and seemingly AI-generated profile pictures, it seems challenging to determine who actually runs these services.

Our work is in its early stages. We analysed just below 3,000 papers, roughly 1,000 of which seem to have been used for citation boosting, while the other papers seem to have been used to support SSPAs. The limited data-size stems from the lack of an effective method for extracting information from ResearchGate at scale and from the limited timeframe of this study. We also restricted ourselves to metadata curated by ResearchGate and did not extract information from the PDF files uploaded to the platform.  To enable collaboration and further research, we are making the dataset used in this study publicly available~\cite{rg}. In addition to addressing these limitations, we plan to also quantify the extent to which fake papers are cited by legitimate authors, an extremely concerning phenomenon that undermines trust in scientific research.


The practice of manipulating bibliometric indicators may be facilitated by limited user verification on platforms like ResearchGate. This form of citation farming may damage the reputation of academic research and poison other platforms in the process, such as Google Scholar, that indiscriminately index publications on the web. We hope our research will contribute to the body of knowledge that can eventually change such practices or reduce the weight of unreliable bibliometric indicators.

\bibliographystyle{ACM-Reference-Format}
\bibliography{references}

\end{document}